# THE EFFECT OF MECHANICAL STRAIN ON PROPERTIES OF LUBRICATED TABLETS COMPACTED AT DIFFERENT PRESSURES

*By*

Pallavi Pawar (1), Hee Joo (1), Gerardo Callegari (1), German Drazer (2), Alberto M. Cuitino (2), and Fernando J. Muzzio (1)*

(1) Department of Chemical and Biochemical Engineering, Rutgers University
(2) Department of Mechanical and Aerospace Engineering, Rutgers University
(*) Corresponding author, 98 Brett Road, Piscataway, NJ 08854. fjmuzzio@yahoo.com 848-445-3357

# Abstract

A full factorial design of experiments was used to study the effect of blend shear strain on the compaction process, relative density and strength of pharmaceutical tablets. The powder blends were subjected to different shear strain levels (integral of shear rate with respect to time) using an ad hoc Couette shear cell. Tablets were compressed at different compaction forces using an instrumented compactor simulator, and compaction curves showing the force-displacement profiles during compaction were obtained. Although the die-fill blend porosity (initial porosity) and the minimum in-die tablet porosity (at maximum compaction) decreased significantly with shear strain, the final tablet porosity was surprisingly independent of shear strain. The increase in the in-die maximum compaction with shear strain was, in fact, compensated during post-compaction relaxation of the tables, which also increased significantly with shear strain. Therefore, tablet porosity alone was not sufficient to predict tablet tensile strength. A decrease in the 'work of compaction' as a function of shear strain, and an increase in the recovered elastic work was observed, which suggested weaker particle-particle bonding as the shear strain increased. For each shear strain level, the Ryskewitch Duckworth equation was a good fit to the tensile strength as a function of tablet porosity, and the obtained asymptotic tensile strength at zero porosity exhibited a 60% reduction as a function of shear strain. This was consistent with a reduced bonding efficiency as the shear strain increased.

# 1. Introduction:

Tablets comprise more than 50% of the total pharmaceutical product units in the worldwide market [1][2], accounting for billions in annual sales. Manufacturing pharmaceutical tablets by direct powder compression involves several processes, including die filling, powder densification, compact formation by inter-particle bonding, and tablet ejection and relaxation. These processing steps can have a significant effect on tablet properties, such as tablet porosity, tensile strength, and dissolution [3]. In fact, our limited understanding of some of these processes, results in loss of yield, product delays and recalls, decreased quality, increased cost, and delayed access to medicine.

The presence of lubricants in the blend can also have a significant effect on tablet properties [4]. It is well known, for example, that tablet tensile strength decreases as lubricant concentration increases. One of the most common lubricants present in pharmaceutical formulations is magnesium stearate (MgSt), and different studies have explained the effect of MgSt concentration on tablet properties. Mollan *et al.* [5] observed that the total work performed during compaction decreased with an increase in lubricant concentration, which was attributed to reduced particle cohesiveness. Wang *et al.* [4] demonstrated the detrimental effect of MgSt on tablet hardness and both Wang *et al.* [4] *and* Uzunović *et al.* [6] found a decrease in dissolution as a function of increasing MgSt concentration. The effect that lubricants have on tablet properties also depends on the excipients present in the formulation and their behavior during compaction [7]. DeBoer *et al.* [8], for example, used SEM imaging to show that MgSt exhibited maximum effect on excipients that undergo plastic deformation without fragmentation and Vromans *et al.* [9] confirmed the effect of MgSt to depend on a number of factors, including particle size, flowability and the mechanism of consolidation of the excipients.

In addition, during tablet manufacturing, the powder blend is exposed to different intensities of shear and normal stress, moisture, and temperature fluctuations. For example, in a typical mixing operation, the particles move relative to the mixer blades and vessel walls as well as relative to each other, leading to the dissipation of mechanical energy via frictional work. The shear experienced by the powder blend can also affect tablet properties and quality, particularly tablet hardness and dissolution. In the presence of MgSt lubricant, it is generally accepted that increased shear leads to an increase in MgSt coating of other excipients. In fact, Pingali *et al.* [10] studied SEM images of powder particles and showed that MgSt coated other materials in the blend as shear increased. Using a modified Couette cell, Mehrotra *et al.* [11] were able to show that the relevant variable to quantify the effect of shear is the induced strain, defined as the integral of shear rate with time, which quantifies the extent of shear, or the *shear degree*. This coating phenomenon also explains the decrease in powder wettability with increasing shear strain, as observed by Pingali *et al.* [12]. As a result, the amount of lubricant and the shear strain experienced by the blend decreases in-vitro dissolution rate [13]. The detrimental effect of excessive shear on the tablet properties mentioned above is generally known as the "over lubrication effect" [14].

If as hypothesized, MgSt is coated onto other particle surfaces, the tablet microstructure could also change, depending on formulation and process parameters. The tablet hardness depends not only on tablet porosity, but is also affected by the bonding between the particles. While there is no direct method to estimate the effect of shear strain on the bonding between the particles, studying the "work of compaction", which involves computing the energy input during different stages of compaction from the compaction curves, has been shown to capture these effects on tablet microstructure. Antikainen *et al.* [15] adopted a similar approach to determine the deformation, fragmentation and elasticity values of different materials. Mollan *et al.* [5] used tablet compaction curves to demonstrate a decrease in total work of compaction with increased lubricant concentration. Osei-Yeboah *et al.* used the concepts of bonding area and bonding strength to explain complex tableting behavior [16]. Tejedor *et al.* discussed the effect of magnesium stearate coating on tablet hardness for plastically deformable and for brittle excipients [17]. These studies attributed the decrease in the hardness to compromised particle-particle bonding in the presence of MgSt.

However, the effect of total shear strain on tensile strength and porosity, both in-die as well as after relaxation, and the effect of shear strain on inter-particle bonding, have not been studied extensively, and are the focus of this work.

## 2. Materials and methods

### 2.1. Materials

In all the experiments, the blend formulation consisted of 90% lactose (w/w), 9% acetaminophen, and 1% magnesium stearate (MgSt). Lactose monohydrate was used as excipient (Foremost farms, Rothschild, Wisconsin), semi-fine acetaminophen as an active ingredient (Mallinckrodt Inc. Raleigh, NC) and MgSt NF as a lubricant (Mallinckrodt, St. Louis, Missouri, lot: P09247). The mean particle size for lactose, acetaminophen and MgSt were 90 µm, 42 µm, and 38 µm, respectively, and they were sieved to remove any possible agglomerates. Acetaminophen is a model drug widely used in industry. It has poor compressibility, compromising the structural integrity of the resulting compacts. Hence, there is a need to study direct compression systems with formulations containing acetaminophen as the active ingredient.

## 2.2 Methods

2.2.1 Experimental Design:

A full factorial design was adopted with 4 levels of shear strain (0, 160, 640 and 2560 revolutions of the Couette device [14], see below) and 5 compaction forces (8, 12, 16, 24, 30 kN).

2.2.2 Blending

First, lactose and acetaminophen were layered from bottom to top, and mixed in a 1.87 L V-Blender (Patterson Kelley) at 15 rpm for 15 min. Then, MgSt was added to this pre-blend and mixed for 25 revolutions. The mixing time for the lubricant was kept low and the blender intensifier bar was not operated in order to have minimum exposure to uncontrolled shear in the blender prior to exposing it to controlled shear in the shearing device (Couette Shear Cell). Two 1kg blends were made and subsequently divided into 300g samples for the modified Couette shear cell.

2.2.3 Shearing- Couette shear cell

There are different ways to impart shear onto blends; varying the mixing time in a blender is one of them. However, this method is scale dependent and powder in different areas in the blender experience different shear rates. Mehrotra *et al.* [11] suggested a systematic approach to expose the blend to uniform shear by using the Couette shear cell (figure 1). A similar approach was adopted in this study. The 300 g samples obtained from the blending procedure were subjected to a controlled uniform shear environment in a modified Couette cell [14]. The Couette cylindrical cell consists of two concentric cylinders, equipped with equidistant baffles inside the annular region. The inner cylinder rotates relative to the outer cylinder, (figure 1). The shear rate was kept constant at 80 rpm for all the samples, except for the first sample which was not sheared in the cell. Three additional samples were subjected to 160 revolutions (sheared for 2 minutes), 640 revolutions (8 minutes), and 2560 revolution (32 minutes). These levels have been shown to span the entire range of shear strain values used in direct compression processes [11].

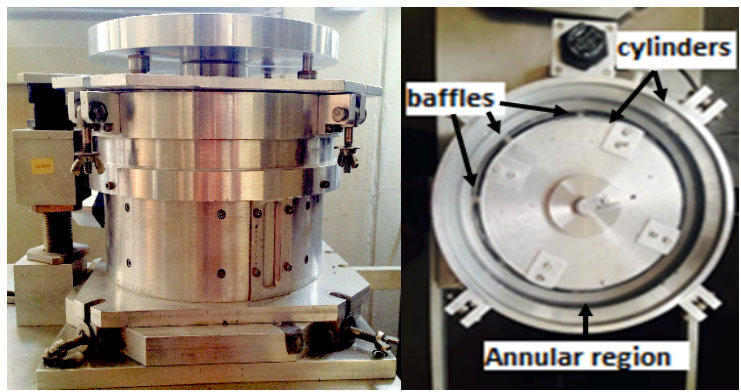

Figure 1: Couette shear cell used to impart uniform shear on blends (left). Top view of the shear cell (right). A 300g aliquot of the blend was poured in the annular region, and the cylinders moved relative to each other imparting the desired level of shear strain.

2.2.4 Tableting

Sheared powder samples were compressed into tablets using a tablet press emulator (Presster$^{TM}$, MCC). The simulated press was a Kikusui Libra2 with both pre-compression and compression rolls of 200 mm diameter. The speed was kept constant and equivalent to 20 RPM in the simulated press. Flat-faced punches and round dies were used to obtain cylindrical tablets of 10 mm diameter. The dosing position was adjusted to obtain 350 mg tablets independent of the shear strain. Tablets were made at a nominal force of 8 kN, 12 kN, 16kN, 24 kN and 30 kN by changing the compaction gap between the punches. The tablet thickness varied from 3.4 mm to 3.0 mm. To reduce tablet variability, the first few tablets were discarded until a stable force and mass were obtained.

2.2.5 Tablet porosity measurements:

The volume of the tablets was calculated from their thickness and diameter. The porosity was calculated from the true density value of the individual components obtained from literature (lactose- 1.54 g/cc, Semifine acetaminophen- 1.29 g/cc, magnesium stearate- 1.03 g/cc) and their respective proportions in a single tablet. The measured true density values from Helium Pycnometry (1.55 g/cc, 1.29g/cc and 1.04g/cc for lactose monohydrate, acetaminophen and magnesium stearate respectively) agreed with those reported in the literature. The blends and tablets were processed and stored in temperature and humidity- controlled environment where the temperature was maintained at $22^0$C and the humidity at 50% RH. None of the materials was known to be hygroscopic and the possible effect of small variations in moisture content on tablet properties was not considered in this study. It was assumed that each tablet contained the targeted concentration of individual components. Tablet porosity was measured both during (in-die) and at the end of the compaction process (out-of-die). The in-die tablet porosity was calculated from the compaction force-thickness profile, obtained directly from the Presster. A certain amount of axial relaxation was observed when the tablets were taken out of the die. The porosity after relaxation was calculated by measuring the tablet thickness after it was removed from the tablet die. The tablet diameter remained constant at 10 mm and comparatively less relaxation (0.4%) was observed in the radial direction compared to the axial relaxation (~ 30%).

2.2.6 Tensile strength testing

The tensile strength was calculated using the Brazilian disk test [18], where the tablet was subjected to diametrical compression between two parallel plates in a Scheulinger tensile strength tester. The tablets cracked in brittle form under tensile stress and the maximum crush force (*F*) was recorded. The tensile strength is then calculated as, $s_{max} = (2*F) / (\pi*D*h)$, where $s_{max}$ is the tensile strength, $F$ is the breaking force, $D$ is the tablet diameter and $h$ is the tablet thickness. We note that the tablets from the condition 2560 rev compressed at 30 kN chipped during the procedure, and hence were not used.

2.2.7 Work of compaction, plastic and elastic energy calculations:

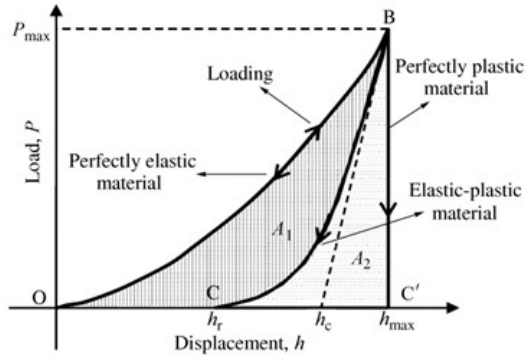

Figure 2: Schematic of a compaction curve. The area under the OB curve (A1 + A2) gives the total work input (Win). The area under the CB curve (A2) gives the elastic recovered work (ERW). Porosity calculated at point O represents the bulk porosity during die fill, porosity calculated at point B represents porosity at maximum compression force and at point C represents final in-die porosity. (Taken with permission from: Iqbal et al., Chin. J. Polym. Sci., vol. 31, no. 8, pp. 1096–1107, May 2013).

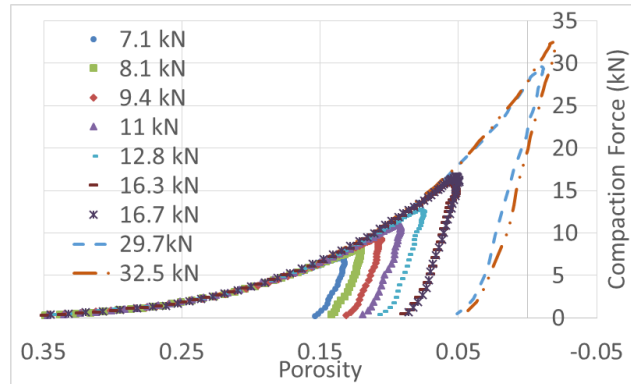

Figure 3: An example of the compaction curves obtained during tableting for different maximum compaction forces. Different colors and symbols correspond to different compaction force.

The work of compaction (total work input, $W_{in}$) was calculated for each tablet from the force-displacement profile (compaction curve) [19]. The total work input ($W_{in}$) during the compaction process is given by area under the OB curve shown in figure 2. As we discussed before, the maximum punch displacement (or minimum tablet thickness) was set to obtain the nominal compaction force, but some small variations in compaction force were present. Note that the height of the peaks in the figure slightly changed with respect to the intended maximum compaction force. During the decompression phase- curve BC in figure 2, the tablet expands and the elastic energy recovered during unloading (elastic relaxation- ERW), is given by the area under the curve. The force-displacement profiles for tablets made at different compaction forces are displayed in figure 3.

## 3. Results and Discussion:

The results are divided into three sections wherein the effect of the process parameters on the porosity and on the tensile strength of the tablets is discussed.

### 3.1 Effect of compaction force and total shear strain on tablet porosity

Figure 4 shows the average porosity of the tablets after relaxation as a function of compaction force for different shear strain levels. Porosity was measured immediately after ejection as indicated in 2.2.5. No further relaxation was observed after a week.

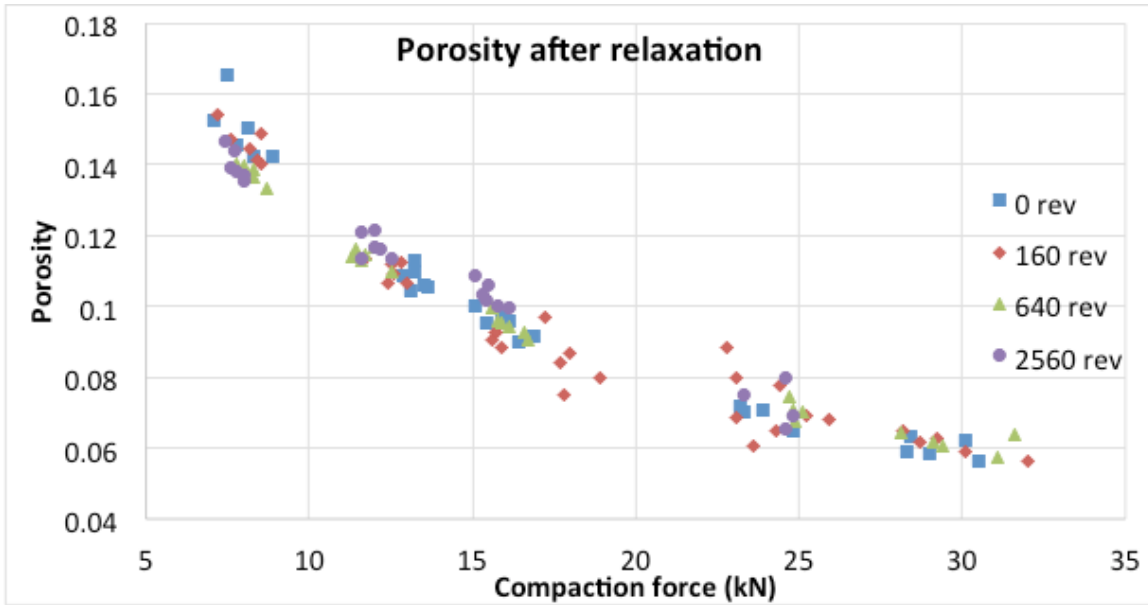

Figure 4: Porosity after relaxation of tablets (out-of-die porosity) compacted at 4 different shear strains and 5 different force conditions. The tablet porosity decreased with an increase in the compaction force, but the blend shear strain had no clear effect.

Tablet porosity decreased with an increase in compaction force, as expected. Surprisingly, however, there was no discernible effect of total shear strain on final tablet porosity. Mehrotra *et al.* observed that the applied shear strain had a significant effect on blend tapped density [11], which could be expected to affect the porosity of the compacted tablets. However, a response surface regression confirmed that only the compaction force had a statistically significant effect on tablet porosity (see table 1). The residuals were normally distributed as shown in figure 5. To validate this conclusion, the shear effect size was taken into consideration [20]. The shear effect size was equal to 2 times the shear regression coefficient. The regression coefficient measures the effect on the mean of the response (here porosity) for a one-unit change in the factor (shear). On the other hand, while dealing with coded units, the effect size for shear is calculated over a two- unit change from -1 (lowest level) to +1 (highest level), and hence is equal to 2 times the regression coefficient. (2*0.003170=0.006340) [21]. The 95% shear confidence limit was computed using the t-statistic. For a 95% confidence interval and error degress of freedom (here =13) , the t-statistic was 2.160. Thus the confidence limit was t-statistic*2*SE coefficient= 0.0078, which was observed to be larger than shear effect. Hence it was concluded that the shear did not have a significant effect for alpha of 0.05.

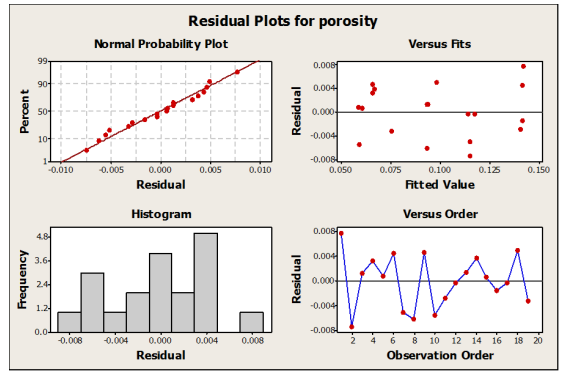

Figure 5: Residual plots for response surface model for porosity. Shear had no significant effect on tablet porosity. The residuals showed no trend with observation order.

```
Response Surface Regression: porosity versus shear, force
The analysis was done using coded units.
Estimated Regression Coefficients for porosity

Term               Coef    SE Coef        T      P
Constant       0.082236   0.003563   23.081  0.000
shear          0.003170   0.001811    1.750  0.104
force         -0.038127   0.002302  -16.564  0.000
shear*shear    0.001919   0.003813    0.503  0.623
force*force    0.019573   0.003128    6.257  0.000
shear*force    0.003364   0.002755    1.221  0.244

S = 0.00499576   PRESS = 0.000825334
R-Sq = 98.09%    R-Sq(pred) = 95.15%   R-Sq(adj) = 97.36%
```

Table 1: Response surface regression to determine the effect of shear and force on porosity after relaxation. Force had a significant effect on tablet porosity. The effect of shear was not significant.

In contrast, the in-die porosity exhibited a clear trend as a function of the shear strain. Figure 6 depicts the evolution of tablet porosity during the compaction process for blends with different shear strain and a nominal compression force of 12 kN. Each curve in figure 6 represents an individual tablet. Similar plots were obtained for the different compaction forces studied here (8, 16, 24, and 30 kN).

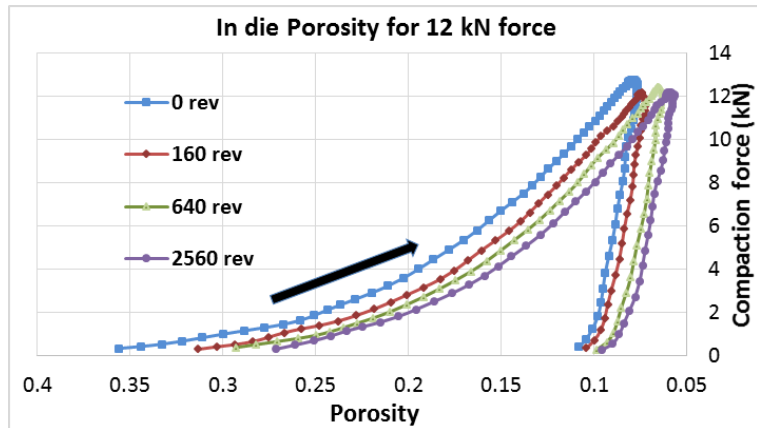

Figure 6: Compaction curves in- die for a compaction force of 12kN. The different symbols correspond to different shear-strain levels as indicated. The bulk porosity during die- fill decreased with an increase in shear strain. Each curve represents a single tablet.

The compaction curves also provide the porosity values at different stages of the compaction process. In figure 7 we plot the average porosity at different compaction stages as a function of shear strain for the case of a nominal compaction force of 12kN. First, the bulk porosity during die filling (corresponding to point O in figure 2) was observed to decrease with shear strain, from 36% to 27%, approximately (figures 6 and 7). This was consistent with previous observations, where the particles exposed to higher shear strain levels packed better during die filling [11]. In addition, the total shear strain had an effect on porosity at maximum compression (point B in figure 2). The corresponding minimum in-die porosity varied from 6% to 9% (figures 6 and 7). The final in-die porosity post unloading (point C in fig. 2), showed less dependence on the shear strain compared to the die-fill porosity, with porosity close to 10%-12% for tablets at all shear strain levels (figures 6 and 7). Thus, a clear convergence in the porosity values for different shear strains was observed, from a maximum difference in the initial bed porosities, to a smaller difference in the minimum porosities at maximum compression and to nearly equal in-die post unloading porosities independent of shear strain. Although most of the axial recovery was observed to occur during the in-die unloading phase, further relaxation was measured after ejection.

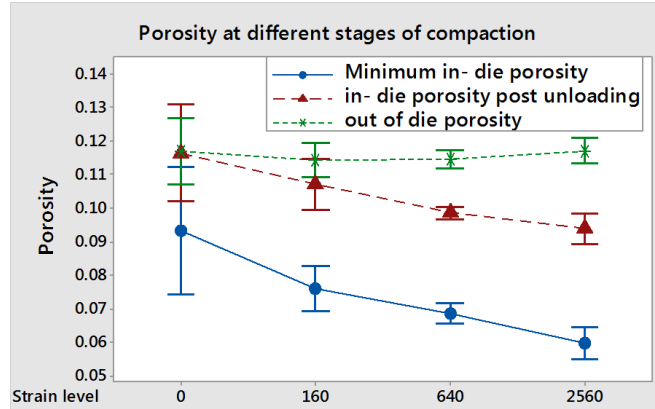

Figure 7: Effect of blend shear strain on porosity at different stages during and post compaction. A part of tablet deformation was recovered during the in- die unloading stage and some additional recovery occurred after the tablet was taken out from the die. The recovery was highest for higher shear strain.

To gauge the effects of shear strain on in-die relaxation during the decompression phase, the apparent axial recovery of each tablet, defined as the increase in porosity during decompression phase with respect to the minimum porosity, was calculated according to the equation:

$$\text{Apparent axial recovery in-die} = (\varepsilon - \varepsilon_{min})*100/\varepsilon_{min} \quad \text{(Equation 1)}$$

where $\varepsilon_{min}$ is the minimum porosity (at maximum compaction) and $\varepsilon$ is the final porosity in- die. The in- die axial recovery during the decompression phase for compacts made at 12 kN force is plotted in figure 8. With an increase in the shear strain level, the tendency of the compacts to relax increased from 28% to 60%. For tablets compacted at 8 kN and 16 kN the axial recovery increased from 16 to 25% and 70 to 147%, respectively. Thus, the tablets compressed under higher compaction forces exhibited larger relaxation.

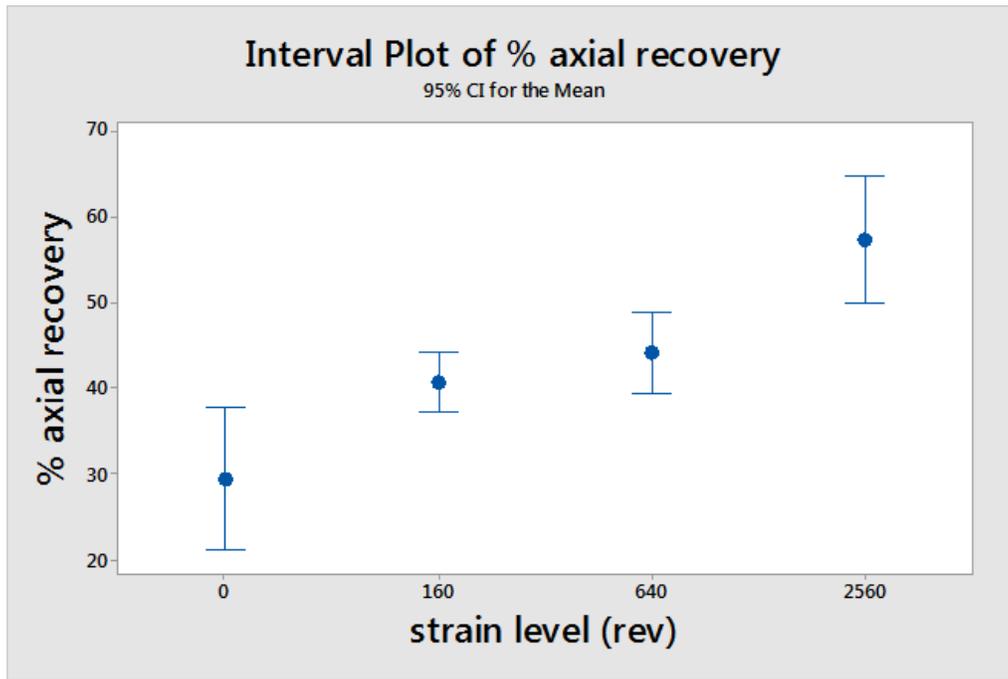

Figure 8: Percentage in-die axial recovery as a function of shear strain for compacts made at a nominal force of 12 kN. The in-die axial recovery increased with an increase in shear strain.

Magnesium Stearate acts as both a lubricant and a glidant, reducing the particle- particle friction as well as the die wall friction, which leads to efficient packing during die-fill. At constant shear strain, tablets showed larger relaxation as a function of the compaction force. More interestingly, relaxation also increased with shear strain, leading to a final porosity that was surprisingly independent of the shear strain.

**3.2 Work of compaction**:

During tablet compaction, the total work input leads to both reversible (elastic deformation) and irreversible processes (plastic deformation, fragmentation, and friction) [22]. A fraction of the elastic energy is recovered (EER) during unloading and ejection, off-die relaxation, and in some cases tablet breakage. The rest remains stored in the tablet in the form of residual stresses, which develop during the loading and unloading process. The ability of the tablet to store residual stresses depends on the formation of bonds during the consolidation process and their strength. Therefore, the elastic energy recovered is inversely related to the bond strength.

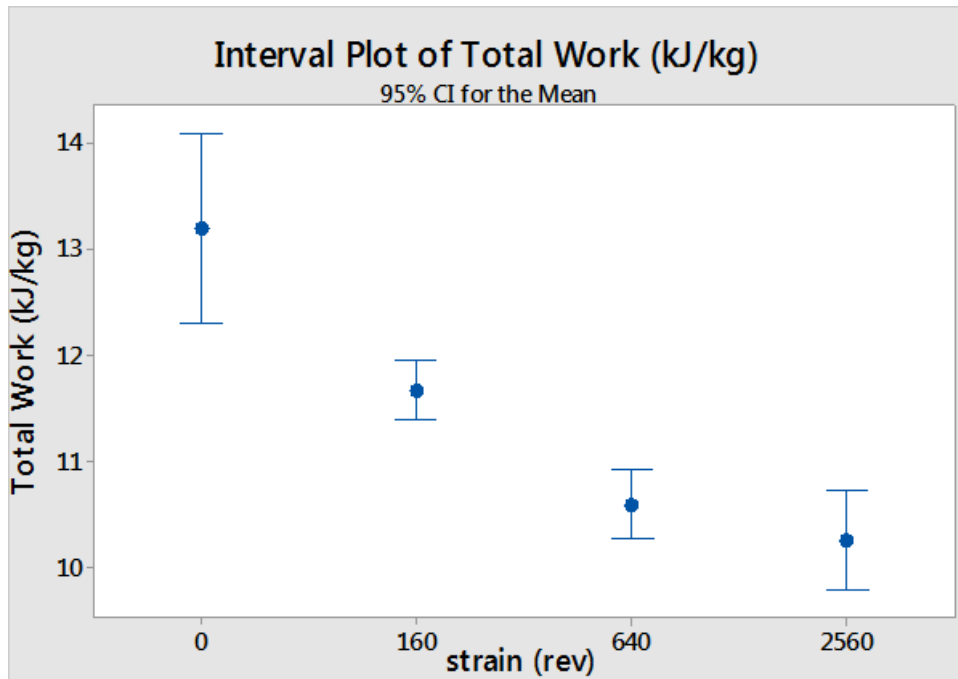

Figure 9: Total work input ($W_{in}$) during compaction (kJ/kg) as a function of total shear strain. The work decreased with increased shear strain.

The total work input increased with the compaction force, as expected, but was observed to decrease with shear strain when the compaction force is constant, as shown in figure 9 for a nominal compaction force of 12 kN. In addition, the elastic energy recovered during the unloading phase per unit of total work input increased with shear strain, as shown in figure 10. This may be attributed to shear strain resulting in weaker particle bonding due to an increase in particle coating with MgSt [23]. Weak compacts resulting from insufficient bonding can only store limited amount of elastic energy in the form of residual stresses after unloading and ejection, leading to a larger fraction of elastic energy recovered. Similarly, during the unloading phase, the axial expansion of the compact increased with shear strain (as seen in figure 8), which was also consistent with weaker bonding.

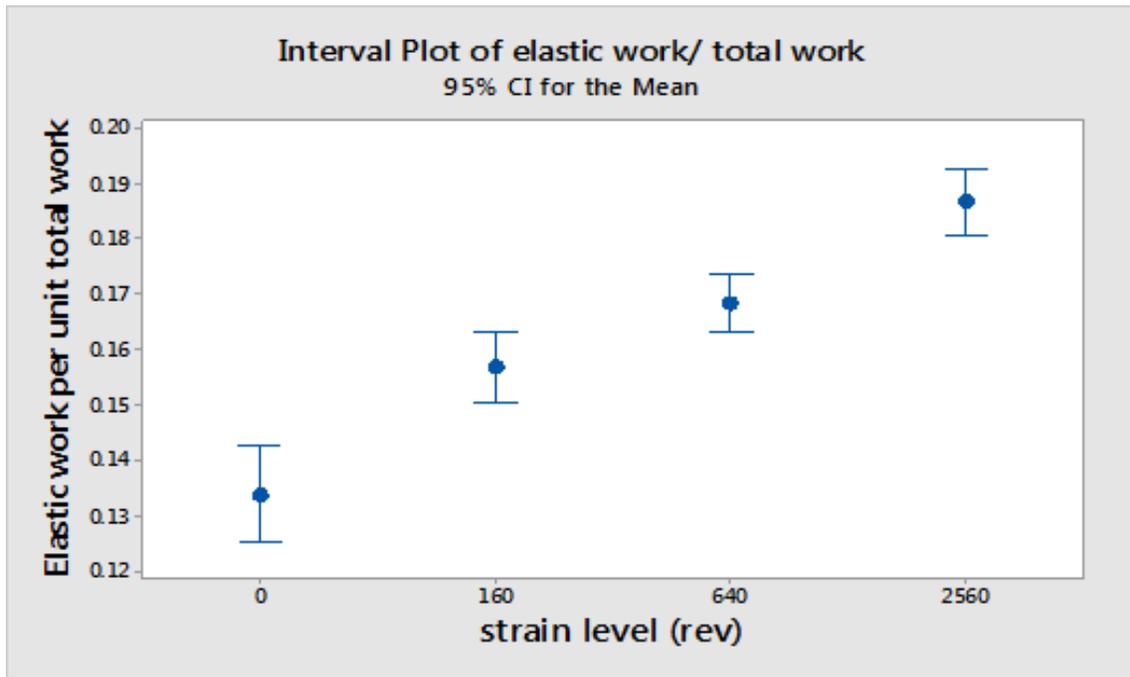

Figure 10: Elastic energy recovered (kJ/kg) during in-die unloading per unit total work as a function of total strain.

**3.3 The effect of blend shear strain and compaction force on tensile strength of tablets**

In figure 11 the measured tensile strength is presented as a function of porosity. As expected, the tensile strength decreased as the porosity of the tablets increased. More important, the tensile strength showed a clear dependence on the shear strain. For a given porosity (or compaction force), the tensile strength of the tablets diminished with increasing shear strain. This confirms that the porosity of the tablets alone was not sufficient to capture the effect of shear strain on the strength of the tablets. In other words, tablets with the same *densification* showed significantly different tensile strengths depending on the shear strain experienced by the blend.

Consistent with our previous observations, this can be attributed to the increased coating of excipients with MgSt at higher shear strains, thus reducing the interparticle contact required for bonding and to create stronger compacts. This is in accordance with Mollan *et al.* reporting that as the total work input during compaction increased, the compacts formed became stronger due to higher amount of energy utilized in the formation of bonds, provided that the die wall friction was minimal [5].

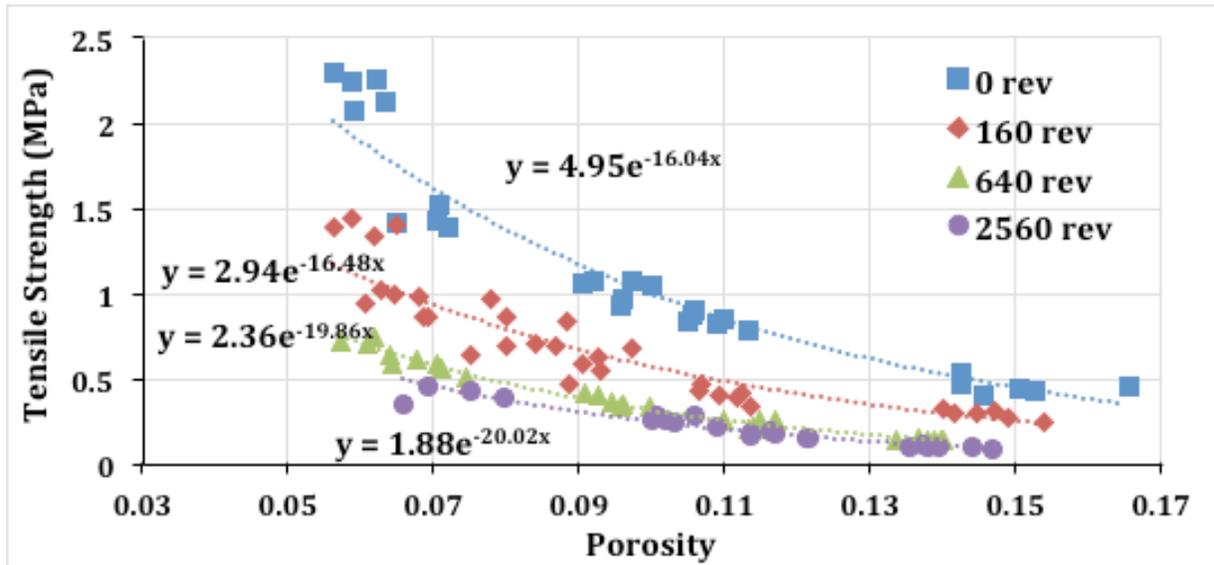

Figure 11: Tensile strength as a function of tablet porosity. The tensile strength was affected by both compaction force and shear strain.

For all shear strain levels, the tensile strength was observed to decay exponentially with porosity (see figure 12 (a)). The tensile strength plots as a function of porosity were fitted to the Ryskewitch Duckworth equation [24]:

$$s = s_0 \exp(-k_s \varepsilon), \qquad (3)$$

where s is the tablet tensile strength, $s_0$ is the tensile strength at zero-porosity and ε is the tablet porosity.

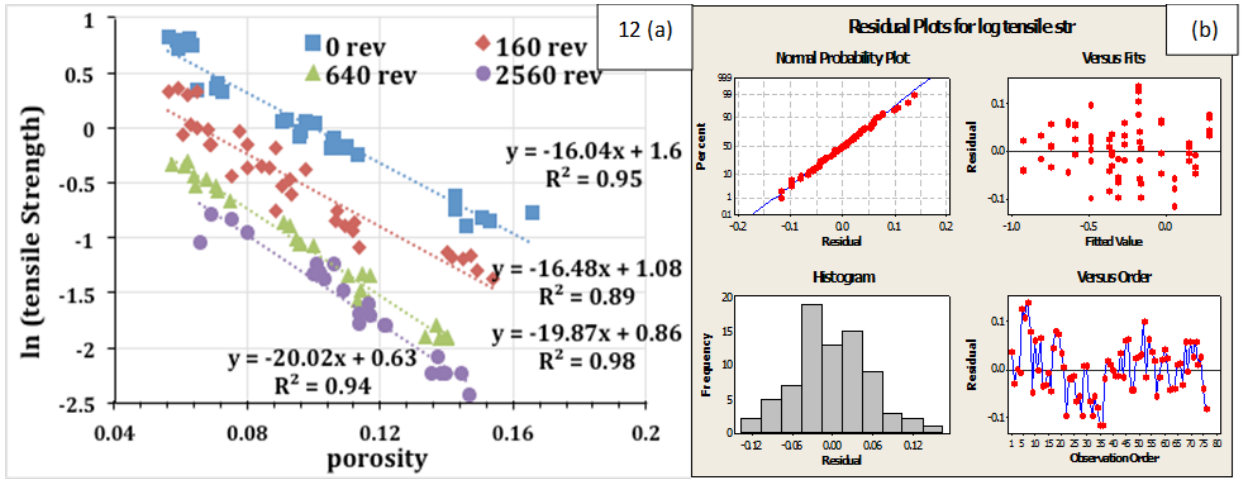

Figure 12: a) Tensile strength of the tablets (in logarithmic scale) as a function of the porosity (left). The dotted lines correspond to the fits obtained for different shear strain levels. b) Residual plots for log tensile strength (right). The residuals were normally distributed.

The fitted parameters are reported in table 2. Interestingly, the effect of shear strain was more significant in the tensile strength at zero-porosity, $s_0$, with a reduction of 60% with shear strain.

| shear strain level | $\ln(s_0)$ | $s_0$ |
|---|---|---|
| 0 | 1.5996 | 4.951052 |
| 160 | 1.0773 | 2.93674 |
| 640 | 0.8567 | 2.355375 |
| 2560 | 0.6311 | 1.879677 |

Table 2: Ryskewitch Duckworth parameters for 12 kN compacts and 4 different strain levels. The tensile strength at zero porosity decreased with an increase in strain.

Finally, a goodness of fit analysis was performed and a response surface was determined by regression to estimate the linear, quadratic and interaction effects of shear and force on the natural logarithm of the tensile strength, leading to an R-square of 97.48% (table 3). The data set involved 80 tablets and the residuals were normally distributed (fig 12 (b)). It was observed that both porosity and compaction force had significant effect on the tensile strength values. Interestingly, the interaction between force and shear appears to be negligible.

```
Response Surface Regression: ln tensile str versus Force, Shear
The analysis was done using coded units.
Estimated Regression Coefficients for ln tensile str

Term               Coef   SE Coef        T       P
Constant       -1.46061   0.04653  -31.391   0.000
Force           0.74229   0.03006   24.694   0.000
Shear          -0.65009   0.02365  -27.484   0.000
Force*Force    -0.24481   0.04085   -5.993   0.000
Shear*Shear     0.97679   0.04980   19.614   0.000
Force*Shear     0.01134   0.03598    0.315   0.754

S = 0.130485    PRESS = 1.36752
R-Sq = 97.48%   R-Sq(pred) = 97.10%   R-Sq(adj) = 97.29%
```

Table 3: Goodness of fit analysis for response surface regression. Both strain and for had a significant effect on the tensile strength.

## 4. Conclusions:

Designed experiments were performed to explore the effect of strain on tablet porosity and hardness. It was concluded that porosity alone was not sufficient to predict tablet tensile strength and that the knowledge of shear strain during processing was also critical. To determine the effect of shear strain, different tablet properties such as tablet porosity during and post compaction, the axial recovery of tablets, the total work of compaction and the elastic work that was recovered during the compaction process were measured.

First, it was shown that, although shear strain had a significant effect on the initial in-die bed porosity, a smaller effect is observed on the porosity at maximum compaction, and no significant effect was present on porosity after relaxation. Most of the tablet axial recovery occurred during the in-die unloading phase and some additional recovery was observed after ejection. The calculated percentage axial recovery also highlighted the differences in relaxation depending on the shear strain levels.

Materials exposed to different shear levels also exhibited different tendencies to bond, which affected both the total work input and the recovered elastic energy. The portion of elastic energy recovered during the in-die tablet relaxation showed an increase with increasing shear strain level. All these observations are consistent with the hypothesis that shear strain induces MgSt to coat the particle surface of other excipients, thus reducing bonding and ultimately affecting the tensile strength of the tablets. In fact, tensile strength was shown to decrease with shear strain. In addition, the tensile strength showed an exponential decay with tablet porosity and was accurately described by the Ryskewitch Duckworth equation. Consistent with our previous results and the hypothesis of MgSt increasingly coating the excipient particles, the strength at zero-porosity obtained from fitting the experimental data showed a 60% reduction with shear strain.

While the results obtained here examined a wide range of shear strain values, only one formulation was explored, and the effect of shear rate was not examined. Effects of ingredient material properties, shear rates, and tablet density on dissolution performance, will be explored in future communications.

**Acknowledgements**

The authors wish to acknowledge the support of the National Science Foundation under Grant number IIP-1237873, Industry-Academia Research Partnership for Developing & Implementing Non-Destructive Characterization and Assessment of Pharmaceutical Oral Dosages in Continuous Manufacturing. The authors would like to thank Dr. Ron Snee for his help with statistical analysis presented in this work.

# 5. References:

[1] "Manufacturing of Solid Dosage Forms - Transformation in Manufacturing Concepts and Expanding Contract Manufacturers' Production Capacity May Force Regulatory Bodies to Re-Design Guidelines on Quality Standards - RnR Market Research." [Online]. Available: http://www.rnrmarketresearch.com/manufacturing-of-solid-dosage-forms-transformation-in-manufacturing-concepts-and-expanding-contract-manufacturers-production-capacity-may-force-regulatory-bodies-to-re-design-guidelines-on-quality-market-report.html.

[2] "Solid Dose Realities: Reaching for More," 2014. [Online]. Available: http://www.pharmamanufacturing.com/articles/2014/solid-dose-realities-reaching-for-more/. [Accessed: 25-May-2015].

[3] C. K. Tye, C. (Calvin) Sun, and G. E. Amidon, "Evaluation of the effects of tableting speed on the relationships between compaction pressure, tablet tensile strength, and tablet solid fraction," *J. Pharm. Sci.*, vol. 94, no. 3, pp. 465–472, 2005.

[4] J. Wang, H. Wen, and D. Desai, "Lubrication in tablet formulations," *Eur. J. Pharm. Biopharm.*, vol. 75, no. 1, pp. 1–15, May 2010.

[5] M. J. Mollan Jr. and M. Çelik, "The effects of lubrication on the compaction and post-compaction properties of directly compressible maltodextrins," *Int. J. Pharm.*, vol. 144, no. 1, pp. 1–9, 1996.

[6] A. Uzunović and E. Vranić, "Effect of magnesium stearate concentration on dissolution properties of ranitidine hydrochloride coated tablets," *Bosn. J. Basic Med. Sci. Udruženje Basičnih Med. Znan. Assoc. Basic Med. Sci.*, vol. 7, no. 3, pp. 279–283, Aug. 2007.

[7] P. J. Sheskey, R. T. Robb, R. D. Moore, and B. M. Boyce, "Effects of Lubricant Level, Method of Mixing, and Duration of Mixing on a Controlled-Release Matrix Tablet Containing Hydroxypropyl Methylcellulose," *Drug Dev. Ind. Pharm.*, vol. 21, no. 19, pp. 2151–2165, Jan. 1995.

[8] A. H. De Boer, G. K. Bolhuis, and C. F. Lerk, "Bonding characteristics by scanning electron microscopy of powders mixed with magnesium stearate," *Powder Technol.*, vol. 20, no. 1, pp. 75–82, May 1978.

[9] H. Vromans, G. K. Bolhuis, and C. F. Lerk, "Magnesium stearate susceptibility of directly compressible materials as an indication of fragmentation properties," *Powder Technol.*, vol. 54, no. 1, pp. 39–44, Jan. 1988.

[10] K. Pingali, R. Mendez, D. Lewis, B. Michniak-Kohn, A. Cuitiño, and F. Muzzio, "Mixing order of glidant and lubricant – Influence on powder and tablet properties," *Int. J. Pharm.*, vol. 409, no. 1–2, pp. 269–277, May 2011.

[11] A. Mehrotra, M. Llusa, A. Faqih, M. Levin, and F. J. Muzzio, "Influence of shear intensity and total shear on properties of blends and tablets of lactose and cellulose lubricated with magnesium stearate," *Int. J. Pharm.*, vol. 336, no. 2, pp. 284–291, May 2007.

[12] K. Pingali, R. Mendez, D. Lewis, B. Michniak-Kohn, A. Cuitiño, and F. Muzzio, "Evaluation of strain-induced hydrophobicity of pharmaceutical blends and its effect on drug release rate under multiple compression conditions," *Drug Dev. Ind. Pharm.*, vol. 37, no. 4, pp. 428–435, Apr. 2011.

[13] H. Abe and M. Otsuka, "Effects of lubricant-mixing time on prolongation of dissolution time and its prediction by measuring near infrared spectra from tablets," *Drug Dev. Ind. Pharm.*, vol. 38, no. 4, pp. 412–419, Apr. 2012.

[14] M. Llusa, *Towards Scientific Manufacturing: The Effects of Shear Rate, Strain, and Composition on the Properties of Blends and Tablets*. ProQuest, 2008.

[15] O. Antikainen and J. Yliruusi, "Determining the compression behaviour of pharmaceutical powders from the force–distance compression profile," *Int. J. Pharm.*, vol. 252, no. 1–2, pp. 253–261, Feb. 2003.

[16] F. Osei-Yeboah, S.-Y. Chang, and C. C. Sun, "A critical Examination of the Phenomenon of Bonding Area - Bonding Strength Interplay in Powder Tableting," *Pharm. Res.*, Jan. 2016.

[17] M. Badal Tejedor, N. Nordgren, M. Schuleit, M. W. Rutland, and A. Millqvist-Fureby, "Tablet mechanics depend on nano and micro scale adhesion, lubrication and structure," *Int. J. Pharm.*, vol. 486, no. 1–2, pp. 315–323, 2015.

[18] H. Mohammed, B. J. Briscoe, and K. G. Pitt, "The interrelationship between the compaction behaviour and the mechanical strength of pure pharmaceutical tablets," *Chem. Eng. Sci.*, vol. 60, no. 14, pp. 3941–3947, Jul. 2005.

[19] T. Iqbal, B. J. Briscoe, S. Yasin, and P. F. Luckham, "Continuous stiffness mode nanoindentation response of poly(methyl methacrylate) surfaces," *Chin. J. Polym. Sci.*, vol. 31, no. 8, pp. 1096–1107, May 2013.

[20] B. Thompson, "Effect sizes, confidence intervals, and confidence intervals for effect sizes," *Psychol. Sch.*, vol. 44, no. 5, pp. 423–432, May 2007.

[21] "Wiley: Design and Analysis of Experiments, 8th Edition - Douglas C. Montgomery." [Online]. Available: http://www.wiley.com/WileyCDA/WileyTitle/productCd-EHEP002024.html. [Accessed: 21-Mar-2016].

[22] S. Patel, A. M. Kaushal, and A. K. Bansal, "Compression physics in the formulation development of tablets," *Crit. Rev. Ther. Drug Carrier Syst.*, vol. 23, no. 1, pp. 1–65, 2006.

[23] A. H. De Boer, G. K. Bolhuis, and C. F. Lerk, "Bonding characteristics by scanning electron microscopy of powders mixed with magnesium stearate," *Powder Technol.*, vol. 20, no. 1, pp. 75–82, May 1978.

[24] "Discussion of Ryshkewitch Paper by Winston Duckworth*," *J. Am. Ceram. Soc.*, vol. 36, no. 2, pp. 68–68, Feb. 1953.